\documentclass{PoS}

\usepackage{amsmath}
\DeclareMathOperator{\lithree}{Li_3}
\DeclareMathOperator{\litwotwo}{Li_{22}}

\title{
Top quark pairs at two loops and Reduze\,2}

\ShortTitle{Top quark pairs at two loops and Reduze\,2}

\author{\speaker{Andreas von Manteuffel}%
         \\
  Institut f\"ur Theoretische Physik,
  Universit\"at Z\"urich,
  8057 Z\"urich, Switzerland\\
  Institut f\"ur Physik (THEP),
  Johannes Gutenberg-Universit\"at,
  55099 Mainz, Germany\\
  E-mail: \email{manteuffel@uni-mainz.de}}

\author{Cedric Studerus\\
  Fakult\"at f\"ur Physik,
  Universit\"at Bielefeld,
  33501 Bielefeld, Germany\\
  E-mail: \email{cedric@physik.uni-bielefeld.de}}

\abstract{We report on progress for the analytical calculation of
the two--loop corrections to top quark pair production at hadron colliders.
For the light fermionic corrections in the gluon channel,
we discuss the analytical solution for the master integrals of a
non--planar double box with a massive propagator.
The result in terms of Goncharov's multiple polylogarithms is handled using
systematic reductions based on the symbol map and the coproduct.
We discuss new features of the computer program Reduze\,2.
It provides a fully distributed variant of Laporta's algorithm to
reduce loop integrals.
New graph matroid based algorithms allow to calculate shift relations
between Feynman integrals in a fully automated way.
}

\FullConference{Loops and Legs in Quantum Field Theory\\
11th DESY Workshop on Elementary Particle Physics \\
April 15-20, 2012\\
Wernigerode, Germany}

\begin{document}

\section{Top quark pair production at NNLO}

Top quark pair production has been studied in detail both at the Tevatron
and the LHC.
In order to match the experimental precision, theoretical predictions beyond
the next-to-leading order (NLO) are necessary~\cite{Bonciani:2011zza}.
Several precision predictions for the total cross section
and some distributions are available~\cite{Beneke:2012wb,Moch:2012mk,Cacciari:2011hy,Ahrens:2011px,Kidonakis:2010dk},
taking into account different approximations
to the next-to-next-to-leading-order (NNLO) corrections and resummations of
logarithmic terms.
For the quark initiated channels, an exact NNLO prediction for the total cross
section was presented~\cite{Baernreuther:2012ws,Czakon:2012zr,Mitov:proc}
employing semi-numerical methods.

For both, the quark and the gluon initiated channels, different building blocks
of the NNLO prediction are known also analytically.
For the two--loop corrections, this includes the high-energy
limit~\cite{Czakon:2007ej,Czakon:2007wk}
and the infrared poles with full mass dependence~\cite{Ferroglia:2009ii}.
In~\cite{quarkloops,planarqq,Bonciani:2010mn}, the fermionic and leading colour
contributions in the $q\bar{q}$ channel and the leading colour contributions
in the $gg$ channel were given.
The one-loop squared contributions are
known~\cite{Korner:2005rg,Korner:2008bn,Anastasiou:2008vd,Kniehl:2008fd}
for some time.
Subtraction terms needed to combine the results with the contributions
for one or two additional partons in the final state were presented in
\cite{Bierenbaum:2011gg,Bierenbaum:proc,Bernreuther:2011jt,Abelof:2011ap,Abelof:2012rv,Abelof:proc}.

Our setup for computing the analytical two-loop corrections to top quark
pair production is as follows.
We generate diagrams with {\tt QGRAF}~\cite{qgraf}, reduce the loop
integrals and compute the interference terms with
{\tt Reduze\,2}~\cite{Studerus:2009ye,vonManteuffel:2012np}.
For the master integrals, we employ the method of differential
equations~\cite{Kotikov:1990kg}.
Integration constants are fixed by evaluating Mellin-Barnes representations
in different kinematical limits.
For planar topologies we use {\tt AMBRE}~\cite{Gluza:2007rt} for the generation
of Mellin-Barnes representations and {\tt MB.m}~\cite{Czakon:2005rk}
for their expansions.

\section{Light fermionic two--loop corrections to $gg\to t\bar{t}$}

\begin{figure}
\centerline{\includegraphics[width=0.4\textwidth]{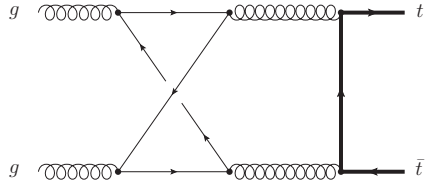}}
\caption{\label{fig:npbox}
A non--planar double box diagram with a massive propagator (thick line)
contributing to the light fermionic corrections to top quark pair production
in the gluon channel.
}
\end{figure}

For the two--loop corrections to $gg \to t\bar{t}$
which involve a closed light fermion loop~\cite{lightfermionamp},
we computed 11 new master integrals in terms of Goncharov's multiple
polylogarithms (GPLs)~\cite{Goncharov:gpl} via the method of differential equations.
The most involved topology is a non--planar double box with one massive propagator,
see figure~\ref{fig:npbox}.
A major source of complexity originates from the fact that this non--planar
topology has cuts in the Mandelstam variables $s$, $t$ and $u$ at the same time.
We find three master integrals for this topology.
For the kinematic variables $x=(\beta-1)/(\beta+1)$ and
$y=-t/m^2$ and a suitable choice of master integrals, the
differential equations decouple after expansion in $\varepsilon=(4-d)/2$ and can be
integrated.
Here, $\beta=\sqrt{1-4m^2/s}$, $m$ is the top mass and $d$ the number of
space-time dimensions.
We fix the integration constants from symmetry conditions as well as
regularity constraints and asymptotic expressions obtained from an explicit
Mellin-Barnes calculation.
The Mellin-Barnes representation was obtained (for a different choice of master
integrals) by direct integration of Feynman parameters similar to the massless
case~\cite{Tausk:1999vh}.
Our Mellin-Barnes setup was successfully cross-checked against
numerical samples obtained with
{\tt SecDec\,1}~\cite{Carter:2010hi,Borowka:2012yc,Heinrich:proc}
for the leading poles.
For the result we employ GPLs defined recursively by
$G(\{w_1,w_2,\ldots,w_n\},z) = \int_0^z G(\{w_2,\ldots,w_n\}, z') \mathrm{d}z'/(z'-w_1)$
with the special rules $G(\{\},z)=1$ and
$G(\{w_1,\ldots,w_n\},z)=\ln^n z/n!$ for $w_1=\ldots=w_n=0$,
see also~\cite{Ablinger:2011te,Bluemlein:proc}.
Our result for the master integrals up to the finite pieces
involves $\mathcal{O}(10^3)$ GPLs with maximal weight 4.
These functions are $G(\{w_1,\ldots,w_{n\leq4}\},y)$ for
$w_i \in \{0,-1,-x,-1/x,-1/x-x,-1/x-x+1\}$ and
$G(\{w_1,\ldots,w_{n\leq4}\},x)$ for
$w_i \in \{0,\pm 1, \pm i, (1\pm i \sqrt{3})/2\}$.
~

The complexity of these multiple polylogarithms makes it difficult to handle
them with traditional methods.
A fast and reliable numerical evaluation via real valued functions,
expansions for asymptotic kinematics and establishing special features such
as symmetries requires a systematic approach to functional identities between
multiple polylogarithms.
The \emph{symbol map} associates to each multiple polylogarithm a linear combination
of tensors which fulfil a "generalised logarithm law" in form of simple
algebraic rules.
It was demonstrated~\cite{Goncharov:2010jf} in the context of $N=4$
Super-Yang-Mills theory that the symbol captures essential information
about the multiple polylogarithms in a form suitable to allow for drastic simplifications of amplitudes,
see also~\cite{Duhr:proc}.
We successfully applied symbol map techniques to our results in the context
of QCD with a mass and were able to considerably simplify expressions for poles
in $\varepsilon$.
For example, the $1/\varepsilon$ pole of the corner integral for the
non--planar topology discussed above contained 39 GPLs with weight 3, while
the simplified expression contains only one weight 3 classical
polylogarithm:  $\lithree((y_1+z_1)/(y_1 z_1))$ with
$y_1=y+1$, $z_1=z+1$ and $z=-u/m^2$.
Recently, algorithmic approaches have been proposed in \cite{Duhr:2011zq}
for the symbol calculus.

A breakthrough for the domain of applicability of these methods has been
achieved by extending~\cite{Duhr:2012fh} the symbol calculus employing the
\emph{coproduct}~\cite{Goncharov:coproduct}.
In this way, also subleading degree terms can be treated algebraically except
for pure constants, which may be addressed by numerically assisted methods.
This was demonstrated in \cite{Duhr:2012fh} for multi--scale two--loop QCD
corrections.
We successfully apply these methods to algorithmically reduce our results
for top quark pair production to a set of independent functions.
Our results show that also for the finite terms of the $\varepsilon$
expansion significant simplifications can be achieved, although not necessarily
as dramatic as for the $1/\varepsilon$ contribution described above.
In contrast to the QCD application in \cite{Duhr:2012fh}, our results contain
genuine multiple polylogarithms, which can't be expressed in terms
of classical polylogarithms alone but require functions such as
e.g.\ $\litwotwo(u,v)$.

It will be interesting to see to what extend symbol calculus based methods can
be used directly for \emph{solving} master integrals in the multivariate case,
see for example~\cite{Bogner:proc,Chavez:2012kn}.

\section{Reduze\,2}

Reduze is a computer program written in C++ to reduce Feynman integrals.
It represents a central tool for our calculations of two--loop corrections
to top quark pair production.
Version\,2~\cite{vonManteuffel:2012np} provides a major rewrite and extension
of its predecessor Reduze\,1~\cite{Studerus:2009ye}.
New features include the distributed reduction of \emph{single topologies}
on multiple computers or processor cores via a distributed variant of Laporta's
algorithm~\cite{Tkachov:1981wb,Chetyrkin:1981qh,laporta}.
The parallel reduction of \emph{different topologies} is supported via a
modular, load balancing job system.
We observe significant speed-ups for using up to $\mathcal{O}(100)$ cores
in applications for $t\bar{t}$ production.

Reduze\,2 also provides fast graph and matroid based algorithms, which allow
for the identification of equivalent topologies and integrals.
These algorithms automatically detect and explicitly construct shifts of loop
momenta such that one set of propagators is transformed into another set
of propagators.
Where applicable, this transformation may be supplemented by a crossing of
external legs.
These \emph{shift relations} can be used to automatically eliminate ambiguities
between loop integrals in the reductions or to map graphs generated by
other programs such as QGRAF to integral families (complete sets of propagators).
Reduze\,2 provides support for multiple integral families and can therefore
essentially work out reductions for full amplitudes in terms of unique master
integrals in a fully automated fashion.
There may be shift symmetries which map a sector
(selection of propagators from an integral family) but not every integral of
it onto itself.
We wish to point out, that in some cases these symmetries provide additional
reductions which are not found by typical finite sets of integration-by-parts
relations.

We propose a new shift finding algorithm to find a possible shift relation
between the \emph{propagators} of two graphs via a \emph{matroid isomorphism}
test.
The latter is reduced to graph isomorphism tests for twisted graphs.

\begin{figure}
\centerline{
\includegraphics[width=0.2\textwidth]{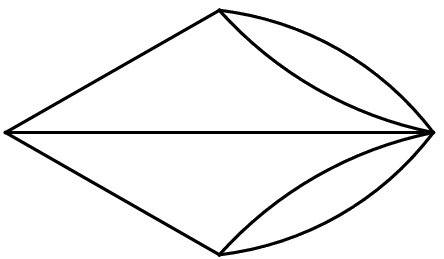}\hspace*{6ex}
\includegraphics[width=0.2\textwidth]{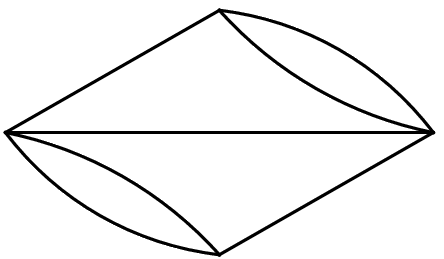}
}
\caption{\label{fig:matroids}
Two vacuum graphs which are not isomorphic but have equivalent
scalar propagators nevertheless.
The graphs are related by a twist and their matroids are isomorphic.
}
\end{figure}

Note that a shift relation may exist even if two graphs are not isomorphic,
see figure~\ref{fig:matroids} for an example.
A more appropriate object to consider is the matroid rather the graph itself.
In this context, matroids generalise the concept of a graph by taking into
account only the linear dependencies of the graph edges without reference
to the vertices.
For simplicity, let us first restrict to biconnected vacuum graphs, where
all propagators have the same mass.
The relevant information of such a graph is encoded in the first Symanzik
polynomial ($\mathcal{U}$ polynomial).
It was observed in \cite{Bogner:2010kv} that two such $\mathcal{U}$
polynomials are isomorphic exactly if their matroids are isomorphic.
In turn, the two graph matroids are isomorphic exactly if the two graphs
are isomorphic after a series of twists.
A twist breaks a graph into pieces by disconnecting edges at two vertices
and reconnecting the subgraphs with flipped orientation.
~

Generalising the idea to cases with external momenta and different internal
masses we end up with the following algorithm to find a possible shift
between two sets of propagators:
\begin{enumerate}
\item Generate graphs for each of the two sets of propagators.
\item Colour the edges according to the masses (prolong lines by introducing
two--point vertices).
\item For each graph, connect the external legs with a new vertex.
\item Decompose each graph into triconnected components to find
all possible twists.
\item Minimise each graph by performing twists
(remove auxiliary vertices from twisted graph and map to canonical label
accounting for graph isomorphisms).
\item Check for graph isomorphism between minimised graphs.
\end{enumerate}
In our tests, this method outperforms a direct combinatorial matching of
propagators by orders of magnitude and is thus applicable to higher
loop orders.

\section{Conclusion}

Considerable progress was made towards the complete NNLO prediction
for top quark pair production at hadron colliders.
We discussed new ingredients such as an analytical result for
the light fermionic two--loop corrections in the gluon channel.
The techniques and tools employed to obtain them are useful also for other
processes.
The symbol and coproduct calculus for Goncharov's multiple polylogarithms
represents a significant step forward in the accessibility
and automated calculation of analytical results for multi-scale amplitudes
beyond the one-loop approximation.
With Reduze\,2 an open source program is available for the fully distributed
reduction of loop integrals.
Moreover, the package provides new graph and matroid based algorithms to
calculate shift relations for Feynman integrals.

\acknowledgments{
This work was supported in part by the Research Center
{\em Elementary Forces and Mathematical Foundations (EMG)} of the
Johannes-Gutenberg-Universit\"at Mainz, the German Research Foundation (DFG)
and the Schweizer Nationalfonds.
}

\end{document}